\documentclass[10pt,journal,compsoc]{IEEEtran}
\usepackage[utf8]{inputenc}
\usepackage{tabularx}
\usepackage{graphicx,color}
\usepackage{balance}
\usepackage[]{algorithm}
\usepackage[noend]{algpseudocode}
\usepackage{amsmath,url,subcaption}
\usepackage{amsfonts}
\usepackage{graphicx}
\usepackage[table,xcdraw]{xcolor}
\usepackage[normalem]{ulem}
\usepackage{comment}

\useunder{\uline}{\ul}{}

\title{Internet of Behavior (IoB) and Explainable AI Systems for Influencing IoT Behavior}

 \author{
\IEEEauthorblockN{
\textbf{Haya Elayan}, 
\textbf{Moayad Aloqaily}, \textit{Senior Member, IEEE}, 
\textbf{Fakhri Karray}, \textit{Fellow, IEEE},
~\textbf{Mohsen Guizani},~\IEEEmembership{Fellow, IEEE}%
}
\IEEEcompsocitemizethanks{

\IEEEcompsocthanksitem \textit{H. Elayan} is with xAnalytics Inc., Ottawa, ON, Canada. \protect E-mail: hh.elayan@xanalytics.ca%
\IEEEcompsocthanksitem \textit{M. Aloqaily, M. Guizani, and F. Karray} are with the Mohamed Bin Zayed University of Artificial Intelligence (MBZUAI), UAE. \protect E-mails: \{maloqaily; mguizani, fkarray\}@mbzuai.ac.ae
}}

\newcommand{\remove}[1]{}

\begin{document}

\maketitle
\begin{abstract}
Pandemics and natural disasters over the years have changed the behavior of people, which has had a tremendous impact on all life aspects. With the technologies available in each era, governments, organizations, and companies have used these technologies to track, control, and influence the behavior of individuals for a benefit. Nowadays, the use of the Internet of Things (IoT), cloud computing, and artificial intelligence (AI) have made it easier to track and change the behavior of users through changing IoT behavior. This article introduces and discusses the concept of the Internet of Behavior (IoB) and its integration with Explainable AI (XAI) techniques to provide trusted and evident experience in the process of changing IoT behavior to ultimately \textcolor{black}{improve user} behavior. Therefore, a system based on IoB and XAI has been proposed in a use case scenario of electrical power consumption that aims to influence user consuming behavior to reduce power consumption and cost. The scenario results showed a decrease of 522.2 kW of active power when compared to original consumption over a 200-hours period. It also showed a total power cost saving of €95.04 for the same period. Moreover, decreasing the global active power will reduce the \textcolor{black}{global} intensity through the positive correlation.
\end{abstract}


\section{Introduction}
Natural disasters and epidemics have affected people's lives for centuries. \textcolor{black}{Tsunami, Spanish flu, HIV/AIDS, the COVID-19 pandemic, among others, have killed millions around the world and forced people to change their behaviours}.
Its impact on increasing mortality \textcolor{black}{rate}, causing economic damage, changing the behavior of individuals, causing social and economic disruption, and increasing political pressure and tensions which compels developed countries to analyze them carefully.

While changing behavior appears to be a low-risk effect compared to mortality or economic disaster, it is powerful enough to cause such risks. For example, in a study of the \textcolor{black}{impact} of social distancing on suicide rates apart from epidemiological deaths, results showed that a change in the social distancing index of ten units caused a 2.9\% increase in the suicide rate \cite{stack2021social}. Moreover, during the COVID-19 pandemic, scientists reported that wearing a mask could prevent approximately 130,000 deaths in the USA alone. 
Many behavioral aspects have been altered by the COVID-19 pandemic, such as customer interaction with brands, employee work procedures, and business engagement with consumers. All these and other examples have economic, technological, and physiological \textcolor{black}{effects}. 
Therefore, tracking people's behavior becomes crucial to influence it in adverse situations. For example, using machine learning for mask recognition tasks is one way to get individuals to respect regulations and monitor negligence. 

By 2023, \textcolor{black}{it} is predicted that the activities of 40\% of the global population will be tracked digitally to influence human behavior \cite{granter}. 
It is also projected that the IoT sensors market will reach \$22.48 billion, and 29.3 billion \textcolor{black}{Internet}-connected devices will be available by 2023. Therefore, IoT would play a major role in detecting people's behavior 

The Professor of Psychology, Gote Nyman, was the first to announce the concept of the Internet of Behavior (IoB). In 2012, he said that if the human behavioral pattern is assigned to devices (e.g. IoT devices) with specific addresses, there will be an opportunity to benefit from the knowledge gained by analyzing the history of patterns in many businesses, societal, health, political, and others fields. 
Behavior is a psychological characteristic that can determine a person's willingness to cooperate or collaborate. Despite the other characteristics: Cognition, Emotion, Personality, and Inter-Communication, behavior is responsible for the tendency to act and is highly dependent on the other four criteria. Therefore, focusing on the behavior will allow \textcolor{black}{knowing} how to influence and treat the person.

\textcolor{black}{Attempting to influence and change user behavior is sensitive, as it may encounter resistance and other psychological factors related to comfort and trust. Emerging technologies such as XAI will help provide the user with the required understanding and trust of any system that uses AI models. XAI aims to use methods and techniques to explain to the user what an AI model does and why, thus, giving a better perception of system operations. Therefore, the process of tracking, analyzing, and influencing user behavior will become much simpler.}

This paper integrates the IoB concept with the XAI concept to propose trusted and understandable frameworks that work in \textcolor{black}{user} behavior-changing areas. Accordingly, it proposes \textcolor{black}{an IoB-based system that utilizes XAI} implemented in a use case scenario of household electrical power consumption. \textcolor{black}{The system aims} to change consumer behavior \textcolor{black}{into an eco-friendly behavior} to reduce power consumption and therefore reduce \textcolor{black}{energy} waste and cost. This framework integrates IoT, AI, Data Analytics, Behavioral Science, and XAI techniques to achieve user and business benefits. The contributions of this paper can be summarized as follows:
\begin{itemize}
    \item To the best of our knowledge, this is the first attempt to investigate the concept of IoB, \textcolor{black}{its} workflow and benefits, challenges, and current industrial directions.
    \item Propose a trusted and comprehensible IoB-XAI-based system that attempts to influence and change IoT behavior \textcolor{black}{to achieve benefits for users and businesses}.
    \item Present a use case scenario of electrical power consumption behavior to influence consumer behavior and raise awareness to control and reduce the power consumption process.
    \item Discuss and compare the main characteristics of related work in the power consumption area.
    \item Highlight potential developments and future directions of the proposed system. 
\end{itemize}

\section{\textcolor{black}{Internet of Behavior (IoB): IoT Devices}}
\textcolor{black}{IoB is using devices to collect a massive amount of human behavioral data and turn it into valuable insights to improve user experience and search experience by changing user behavior, interests, and preferences. With the increasing number of IoT devices collecting massive amounts of data, using these devices to track behavioral data in the IoB process will make it easier, faster, and more efficient.}

\begin{figure*}[!htb]
    \includegraphics[width=1\linewidth]{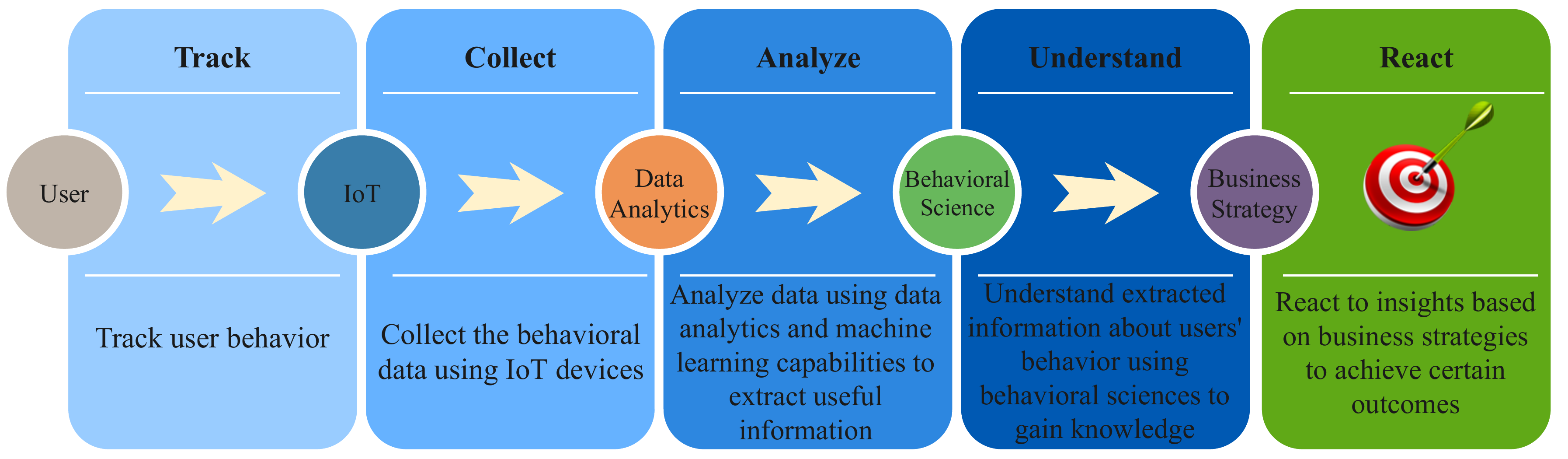}
    \caption{IoB Workflow}
    \label{fig:iob_workflow}
\end{figure*}

IoB attempts to appropriately understand data and use this understanding to create new products, promote current products, redesign the value chain, increase profits, or reduce costs from a psychology perspective.
Therefore, user behavior in the IoB workflow will be tracked first using connected devices, as shown in Fig. \ref{fig:iob_workflow}. The data generated by IoT devices will be collected and then analyzed using data analytics and machine learning algorithms. The analysis phase will yield useful information that must be properly understood from a behavioral science perspective. Finally, the knowledge gained will be used to develop business strategies and influence the behavior of users, thus achieving a specific goal.

\begin{figure}[!hb]
    \centering
    \includegraphics[width=0.8\linewidth]{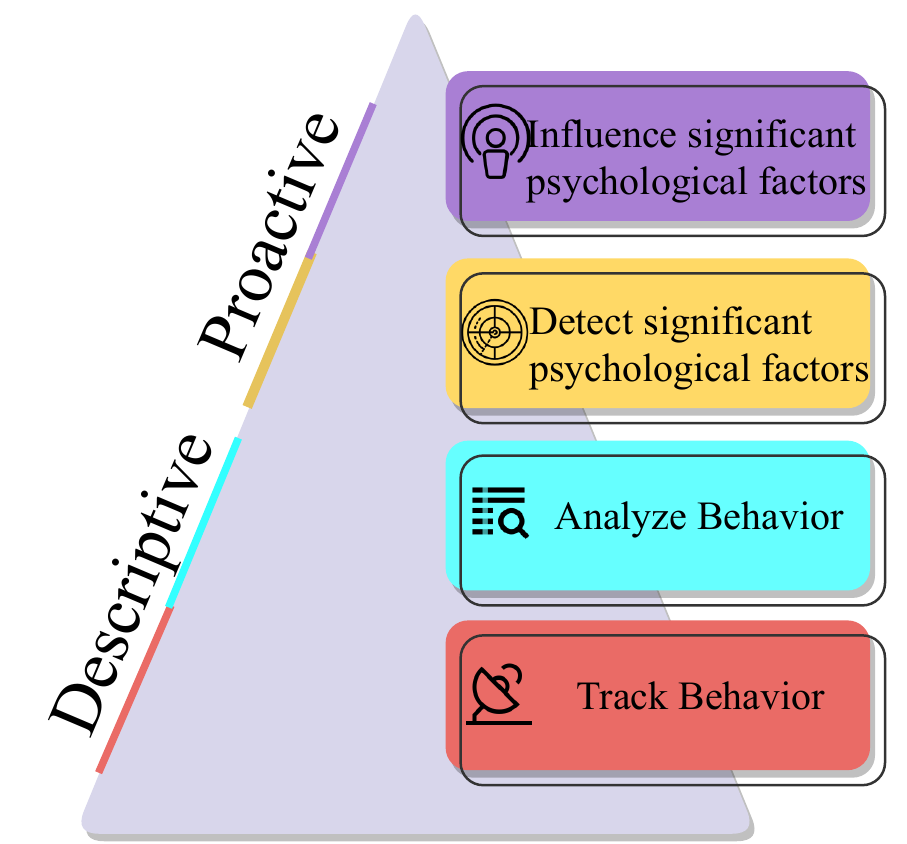}
    \caption{IoB Process}
    \label{fig:iob_process}
\end{figure}

As shown in Fig. \ref{fig:iob_process}, the IoB process is descriptive yet proactive\textcolor{black}{. It} tracks and analyzes user behavior in order to detect and accordingly influence important psychological variables to achieve the process objective. This usually brings lots of benefits as well as a huge set of challenges.


\subsection{IoB Benefits}
IoB is a very new concept in today's emerging technologies. It \textcolor{black}{is} being used in many applications for several benefits, however, without much knowledge about it in the research community. Therefore, in this sub-section, we summarize IoB benefits:
\begin{itemize}
\item \textbf{\textcolor{black}{Quality of Experience}, Increased Profit}: Help companies resolve issues in targeting more sales while keeping their customers satisfied at the same time (Win-Win strategy). \textcolor{black}{For example, online fashion retailer apps can use the clicks and search history of users to suggest personalized discounts and offer packages.}

\item \textbf{Tasks Automation}: Help replace outdated fashion tactics like time-consuming and unfavorable customer surveys (No More Tedious Tasks).

\item \textbf{Target Customers}: Give the opportunity to know the valuable customers based on their interest and which segment to target and invest more time with.
\textcolor{black}{For instance, the smartwatch company may find through tracked data that males aged 20-30 who do not exercise regularly are more likely to purchase its smartwatch to make themselves more committed to exercise.}
\item \textbf{Accuracy}: Give the ability to track and study the unobtainable behavior of how customers interact with products and services (Notice the Unnoticed).

\item \textbf{Real-Time Interaction}: Help provide real-time interactions via notifications or alerts to customers about targeted offers, sales, and/or advertisements. \textcolor{black}{For example, social media apps provide real-time feeds and ads based on viewing patterns, amount of time spent in one account, interest in specific content, or search and chat analysis. }
\end{itemize}

\subsection{IoB Challenges}

\subsubsection{Security}
Dealing with sensitive and real-time data always creates security concerns for any service user. In IoB, it is necessary to demonstrate a high level of security and protection to users, as their behavioral data is expected to be prone to attacks. The data sensitivity motivates cybercriminals to access, reveal, collect and benefit from it.

\subsubsection{Ethical Use}
Behavioral data is a sensitive and personal data type, and its collection, storage, or analysis must be accompanied by transparency and ethical use. \textcolor{black}{Users have} the right to be aware of this process as well as to know that \textcolor{black}{their} privacy is preserved and protected from misuse. \textcolor{black}{Another aspect of ethical use is not following the principle "the end justifies the means". Companies, in most cases, aim for more profit regardless of the impact on users of changing their behavior in critical areas such as health. Lastly, companies must ensure that users' consent is obtained when their data is collected and used for any purpose.}

\subsubsection{Ostrich Effect}
The ostrich effect is a phenomenon that occurs when the rational mind believes something is important and the emotional mind expects it to be painful. People understand the power and benefits of using IoB\textcolor{black}{, however}, they may be uncomfortable (trust issues) about the tracking process, which results in avoiding or rejecting it. Therefore, when companies implement IoB, it is not enough to work on process understanding by users. They also have to work on removing the discomfort of using this technology.

\subsection{XAI for IoB}
IoB challenges may create resistance \textcolor{black}{on the part of users} as it tries to influence and change their behavior and track their data. Using AI models in IoB to analyze users' data allows taking advantage of XAI techniques to make the operations of IoB-based systems more intelligible to the user and thus more trustworthy.

\textcolor{black}{Describing the system process, workflow, impact, and outcomes using user-friendly mathematical and analytical proofs will give users a better understanding of the system and increase awareness of its impact. Thus, \textcolor{black}{mitigating} the problems of trust and the impact of resistance.}

\section{Eye on Industry}
In this section, we will discuss how the different industries are interacting with IoB technology.

\subsection{Healthcare}
Nowadays, smartphones allow many software companies to build and promote different applications that care about \textcolor{black}{user's} health, where they can easily use smartphone sensors to track bio-metrics and healthy behavior. For instance, the \textit{Beddit} health app tracks sleep patterns, heart rate, and breathing rate. After analyzing the tracked behavior, the app sends alerts, notifications, and tips to improve the user's sleep and motivates him/her to achieve daily goals to reach a more positive outcome \cite{Beddit}. 

During the COVID-19 pandemic, many countries have developed health smartphone apps for citizens to control the spread of the virus. The \textit{China Health Code Alipay} app, for example, \textcolor{black}{tracks a user's travel history}, contact history, and body bio-metrics such as temperature. Then, a colored QR code is generated to identify \textcolor{black}{his/her} health status. Accordingly, some restrictions may apply to the user and affect \textcolor{black}{his/her} behavior, such \textcolor{black}{as permitting travel or quarantining at home or in a central location} \cite{alipay}.


\subsection{Transportation}
Even in transportation, IoB can be applied for benefits \textcolor{black}{for drivers and} passengers. After a long history of disputes with drivers and a high turnover rate, Uber is trying to resolve these issues and settle disputes in its favor by influencing driver behavior using gamification. Uber has tricks like loss aversion, recognition, and intrinsic motivation to reward drivers or make them fear losing their winnings \cite{uber}. Also, Ford is expanding its reach by joining forces with Argo AI to develop autonomous vehicles that adapt and behave differently to road infrastructure designs and driving behaviors. This technology is being tested in many locations such as the streets of Miami, Washington, and others \cite{ford}.

\section{IoB and XAI For Power Consumption Reduction: A Use Case Scenario}
\textcolor{black}{This section proposes a use case scenario of a system for household electrical power consumption. The system aims to influence the electricity consumption behavior of a house by automatically controlling the amount of global active power in a way that the user understands and trusts. Global active power control will reduce energy consumption which will, in turn, reduce waste and cost of potential used energy.}

\begin{figure}[!htb]
    \centering
    \includegraphics[width=\linewidth]{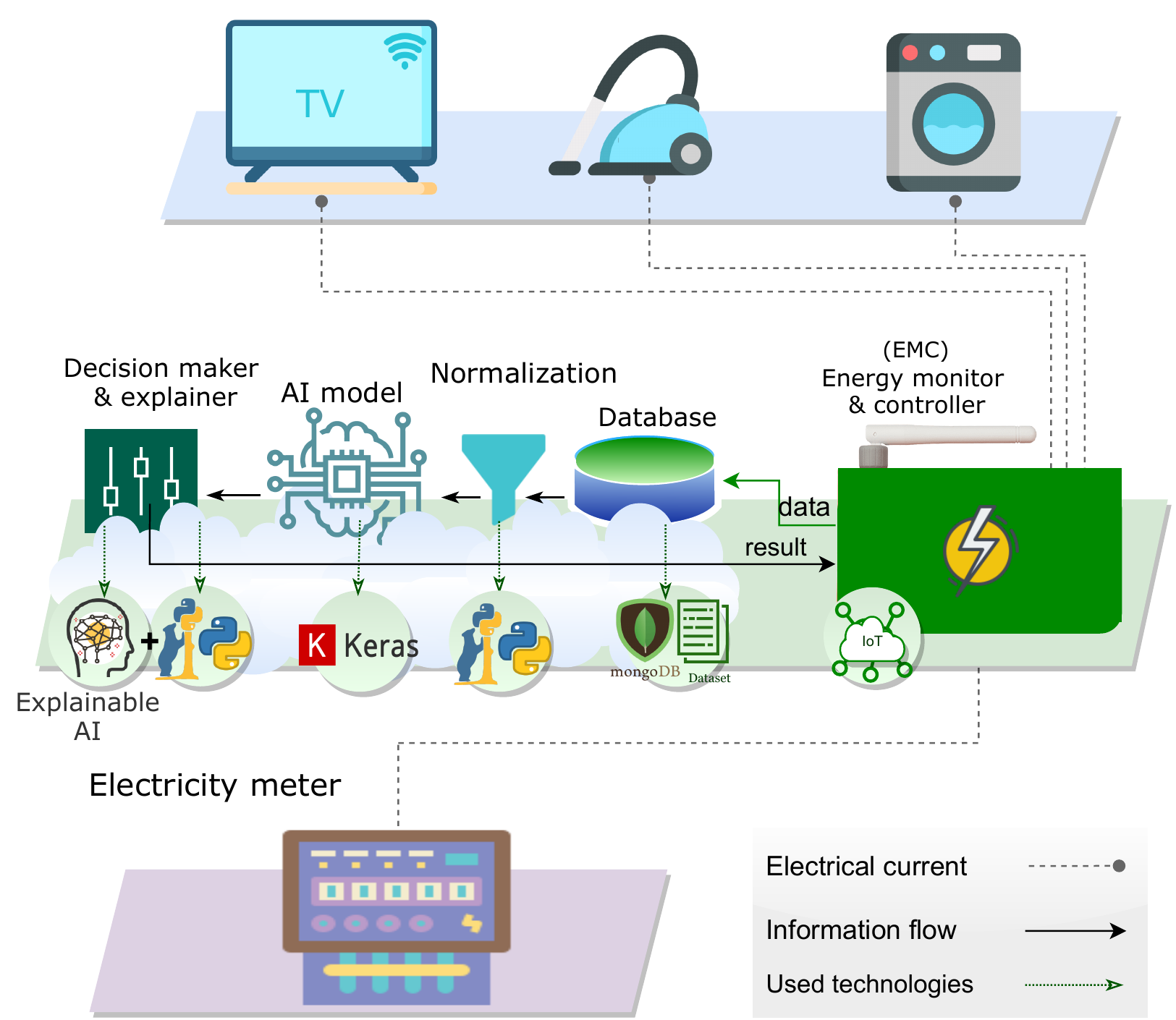}
    \caption{System Workflow}
    \label{fig:system}
\end{figure}

The system workflow in Fig. \ref{fig:system} shows that the current will pass through an energy monitor and controller device (EMC) to know how much electricity each IoT device is using. After that, the data will be transmitted from the EMC to a cloud database to be stored and normalized\textcolor{black}{. Then, the normalized data will be used} by an AI model to predict the global active power that may be consumed in the next hour. The prediction result will be passed to a decision maker and explainer to determine how much energy the appliances should use\textcolor{black}{. Then,} the results will be returned to the EMC to control the consumption in the next hour if it is above a certain threshold in order to reduce it.


Before discussing the proposed system flow in detail, a set of related proposals in \textcolor{black}{the} electrical power consumption area will be discussed. Table \ref{tab:RelatedWork} shows a summary of each related proposal and compared \textcolor{black}{it} to the proposed system (last row in the table).
Studies  \cite{alhussein2020hybrid}, \cite{lim2020deep}, and \cite{wang2021building}, used deep learning algorithms: Long short term memory (LSTM), Convolutional neural network (CNN), or both to predict the electrical consumption. However, they focused their contribution on the architecture of the proposed model. Likewise, \cite{syed2021household} and \cite{kim2020accelerated} used LSTM and CNN together.

The researchers in \cite{syed2021household} have proposed to predict power consumption through proposing a framework consisting of data cleaning and model building steps. Whereas the authors in \cite{kim2020accelerated} aimed to accelerate the deployment process of deep learning models using the proposed edge-cloud framework. 

On the other hand, \cite{liu2020optimization} and \cite{chung2020distributed} applied Deep Reinforcement Learning algorithms (Deep RL) on the same dataset to perform tasks related to process optimization of power consumption. While authors of \cite{liu2020optimization} focused on the proposed RL model structure, authors of \cite{chung2020distributed} proposed a stochastic game framework that controls the interaction between the consumed power and household.

The proposed model of this paper presents a comprehensive framework consisting of all technology needed to form a successful IoB/XAI model which will reduce power consumption by using IoB and XAI to influence user behavior. Firstly, the LSTM algorithm will be used to build a power consumption prediction model, and then the power consumption process will be controlled. Meanwhile, the workflow will be explained to the user to ensure his understanding \textcolor{black}{of} a smoother behavior change process.

\begin{table*}[!htb]
\caption{Related Work Comparison}
\label{tab:RelatedWork}
\resizebox{\textwidth}{!}{%
\begin{tabular}{|l|l|l|l|l|}
\hline
\rowcolor{blue!15}
\textbf{Ref.} & \textbf{Algorithm} & \textbf{Dataset} & \textbf{Goal} & \textbf{Proposal /  Framework description} \\ \hline
\cite{alhussein2020hybrid} & \begin{tabular}[c]{@{}l@{}}CNN\\ LSTM\end{tabular} & \begin{tabular}[c]{@{}l@{}}Short-term individual households\\ power consumption for Australian\\ customers\end{tabular} & Predicting the individual household electrical load. & Focus on deep learning model architecture \\ \hline
\rowcolor{blue!15}
\cite{lim2020deep} & LSTM & \begin{tabular}[c]{@{}l@{}}Monthly electricity consumption\\ from a Korean region\end{tabular} & \begin{tabular}[c]{@{}l@{}}Forecast the annual electricity consumption\\ based on the history of the last three years.
\end{tabular} & Focus on deep learning model architecture \\ \hline 
\cite{wang2021building} & \begin{tabular}[c]{@{}l@{}} CNN \\ ResNet\end{tabular} & \begin{tabular}[c]{@{}l@{}}Electrical meters datasets of non\\residential building in Switzerland\end{tabular} & \begin{tabular}[c]{@{}l@{}}Forecasting  building load using Resnet to enhance\\  the ability of learning\end{tabular} & Focus on deep learning model architecture \\ \hline
\rowcolor{blue!15}
\cite{syed2021household} & \begin{tabular}[c]{@{}l@{}}CNN\\ LSTM\end{tabular} & \begin{tabular}[c]{@{}l@{}}Individual household electric power\\ consumption In France / \\ Appliances Energy Prediction\end{tabular} & \begin{tabular}[c]{@{}l@{}}Predicting electricity consumption at various \\distribution and transmission systems' levels.\end{tabular} & \begin{tabular}[c]{@{}l@{}}Framework of two steps: \\ data cleaning and model building\end{tabular} \\ \hline
\cite{kim2020accelerated}& \begin{tabular}[c]{@{}l@{}}CNN\\ LSTM\end{tabular} & \begin{tabular}[c]{@{}l@{}}Household power consumption\\ by Korea Electric Power Corporation.\end{tabular} & \begin{tabular}[c]{@{}l@{}}Accelerating a deep learning model \\ deployment after power data stream training\end{tabular} & \begin{tabular}[c]{@{}l@{}}Framework for a power flow edge \\cloud system\end{tabular} \\ \hline \rowcolor{blue!15}
\cite{liu2020optimization}& Deep RL & \begin{tabular}[c]{@{}l@{}}Electrical submeters recorded  by\\ Pecan Street. Inc\end{tabular} & \begin{tabular}[c]{@{}l@{}}Optimizing home energy management  by taking\\ a series of actions\end{tabular} & Focus on deep RL architecture \\ \hline
\cite{chung2020distributed}& Deep RL & \begin{tabular}[c]{@{}l@{}}Electrical submeters recorded  by\\Pecan Street. Inc\end{tabular} & \begin{tabular}[c]{@{}l@{}}Scheduling energy consumption for \\household appliances \end{tabular} & \begin{tabular}[c]{@{}l@{}}Framework for the interaction of  power \\ grid and households using a stochastic\\ game \end{tabular} \\ \hline \rowcolor{blue!15}
\begin{tabular}[c]{@{}l@{}}Proposed\\Model\end{tabular} & LSTM & \begin{tabular}[c]{@{}l@{}}Individual household electric power \\ consumption In France\end{tabular} & \begin{tabular}[c]{@{}l@{}}Influencing electricity consumption behavior by\\ automatically controlling the amount of power.\end{tabular} & \begin{tabular}[c]{@{}l@{}}IoB and  XAI based framework of four\\ steps: Data, Database, Data Normalization,\\AI model, Decision making, and XAI\end{tabular} \\ \hline
\end{tabular}%
}
\end{table*}

\subsection{Dataset}
An individual household electric power consumption dataset available online has been used to implement the use case \cite{dataset}. It contains measurements of one house's electrical power consumption per minute for approximately 4 years, collected in a house located in France. \textcolor{black}{The data has been re-sampled to get the total consumption per hour which reduced the dataset size from 2,075,259 to 34,589. The dataset was divided into 50\% train samples and 50\% test samples.} 
\textcolor{black}{The dataset includes the following features: Date, Time, and Voltage, Global active power: Household global active power, Global reactive power: Household global reactive power, Global intensity: Household global current intensity in ampere, Sub-metering 1: Energy of the kitchen, Sub-metering 2: Energy of the laundry room, Sub-metering 3:  Energy of the electric water heater and an air conditioner.}

\subsection{Database and Data Normalization}
Electrical power consumption data is acquired from smart energy monitoring sensors connected to the \textcolor{black}{Internet}. An advanced type of smart monitoring sensor can detect the type of each appliance by connecting the monitor to the electricity meter only,\textcolor{black}{without the need to connect multiple monitoring sensors to each appliance, This helps centralize the system and reduce the cost of its installation.}

The acquired data is huge and in its raw format. Therefore, it \textcolor{black}{requires} a proper database to handle it. 
For example, MongoDB is one of the NoSQL databases that can handle this data as it is document-oriented. \textcolor{black}{MongoDB} can deal with semi-structured data where most IoT data exists in semi-structured or unstructured formats. Moreover, it has high performance in data storage and retrieval, which is necessary for second-by-second data acquisition and transmission. 

Due to the possibility of incomplete, unstructured, and inconsistent data as IoT data is unstructured, the data normalization phase comes to normalize and pre-process the data to get it in a suitable format without noise and outliers. Additionally, this phase handles data loss that may occur due \textcolor{black}{to a disconnect} or dropouts, so it deals with missing values by adding data averaging or removing these values. Moreover, it scales features using scaling methods like Min-Max-Scaler to map data to the same range.

\subsection{Power Consumption Prediction Model}
The dataset used in the scenario is time-series-based data. Time series prediction problems are one of the difficult types of predictive \textcolor{black}{modelling} problems because they add sequence dependence complexity amidst input features. The Long-term memory network (LSTM) is a type of \textcolor{black}{Recurrent Neural Network (RNN)} used in time series problems because it can successfully train huge structures with eliminating major problems, such as vanishing gradients.

A sequential deep learning model has been built to predict the hourly power consumption\textcolor{black}{. I}t allows building a layered model, with each layer having weights corresponding to the following layer. Keras library has been used for coding the model structure. The model has been constructed using one LSTM layer and one Dense layer and ran for 20 epochs with Adam optimizer. The model has been trained on a machine with the following specifications: Intel CPU@2.3GHz, 2 cores, RAM 13GB, and Disk 70GB.

\subsection{Decision Maker and Explainer}
\textcolor{black}{The EMC will track the power consumption of appliances per hour and pass the tracked data to a cloud database to be stored and normalized. The data will be used by the AI model to analyze the consumption behavior and predict the power consumption for the next hour. Based on prediction and historical data patterns, the decision maker part will decide the amount of power that should be used in the next hour}. To personalize the user experience, the decision maker will make the decision based on the \textcolor{black}{historical} user's behavior and usage pattern. Therefore, the decision will be made by comparing the predicted consumption with the \textcolor{black}{historical} average consumption at specific times related to the prediction hour. Take into account influences seasonal and occasional factors.

\textcolor{black}{The explainer part in the Decision Maker and Explainer is responsible for applying the XAI techniques to describe the AI model, its impact, biases, and outcomes to the user. The purpose is to use a set of processes and methods to allow the user to understand and trust the results generated by the model, thereby raising their awareness of the workings of the system and its goal in order to have a smooth behavior change process.}

\textcolor{black}{The explainer will attempt to cover the following concerns: Model prediction process, Model success and failure, and Model confidence in decision making, using data annotation and deep learning algorithm annotation. Data annotation describes statistical information of the dataset, such as features distribution and correlation. While the algorithm annotation shows the characteristics of the model and whether it is biased towards certain features or outliers. The results will be sent to the user whenever they are requested using any \textcolor{black}{Internet}-connected device.}

\textcolor{black}{Using XAI frameworks such as IBM AI Explainability 360 will help provide explanations for various user concerns by providing the dataset and the built model for the open-source toolkit, then selecting the type of consumer to customize the explanations \cite{IBM_XAI}. Finally, form different questions that cover the user concerns and offer explanations and insights through the answers. For example:
\begin{enumerate}
    \item Why is power consumption controlled? The predicted power consumption based on the last hour consumption is 100 kW, your historical average is 55 kW.
\item What appliance consumes the most energy? Appliance "A" is the most consuming device.
\item How was energy consumption distributed during this decision? Appliance “A” consumed 23 Wh, appliance “B” consumed 13 Wh, and appliance “C” consumed 15 Wh.
\end{enumerate}}
Accordingly, the entire system works to influence the user’s consumption behavior in his favor by saving \textcolor{black}{energy} from waste, saving the cost of wasted \textcolor{black}{energy}, and increasing \textcolor{black}{his/her} awareness of better consuming behavior. Therefore, creating `'collective wisdom" and more responsible habits that affect not only the individual but also the society as a whole.

\subsection{Use Case Results}
\subsubsection{\textcolor{black}{Model Performance}}
\begin{figure}[!hb]
  \subfloat[Model train vs. test loss]{\includegraphics[width=0.45\textwidth]{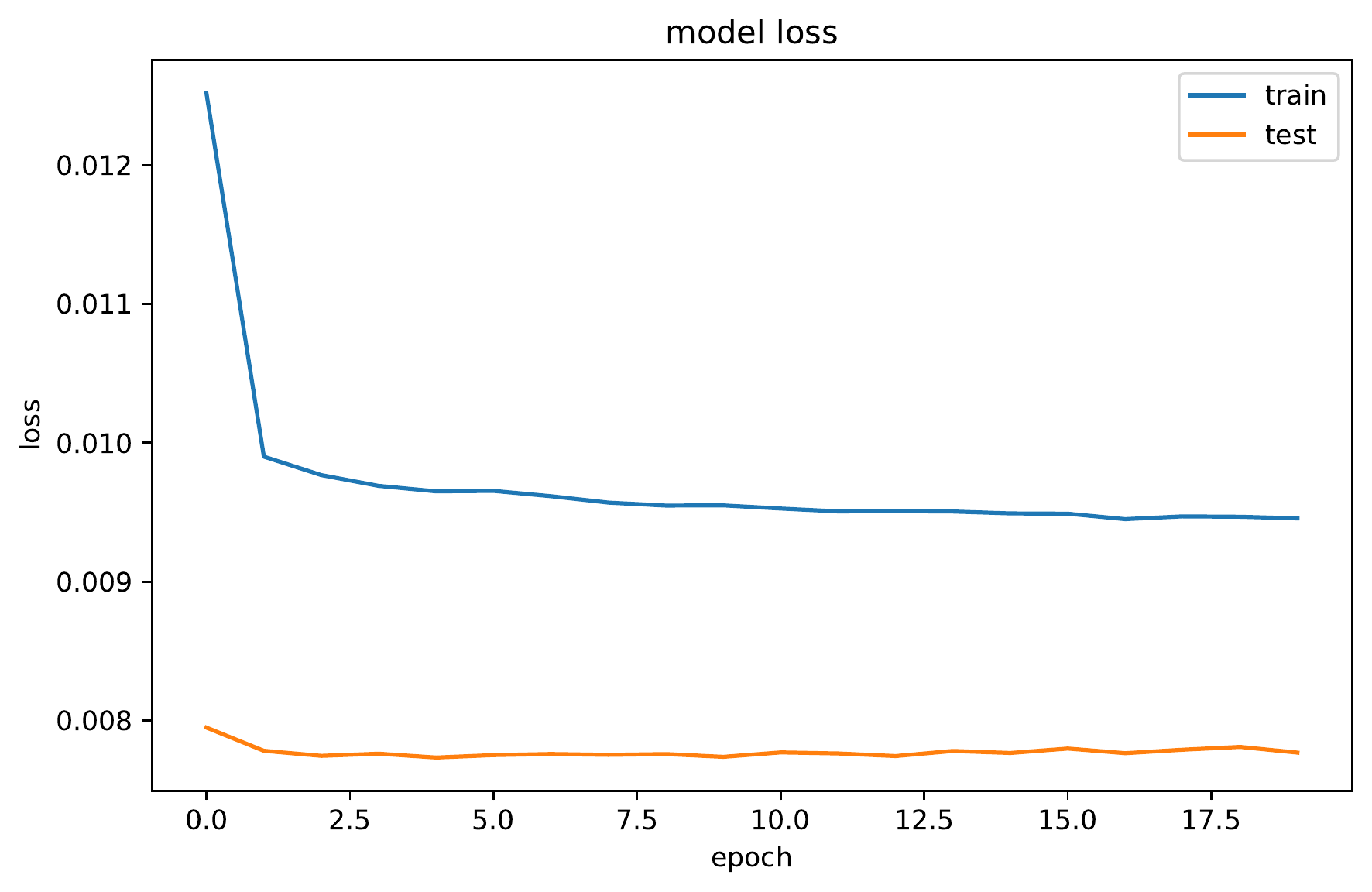}\label{fig:loss}}
  \hfill
  \subfloat[Model results]{\includegraphics[width=0.45\textwidth]{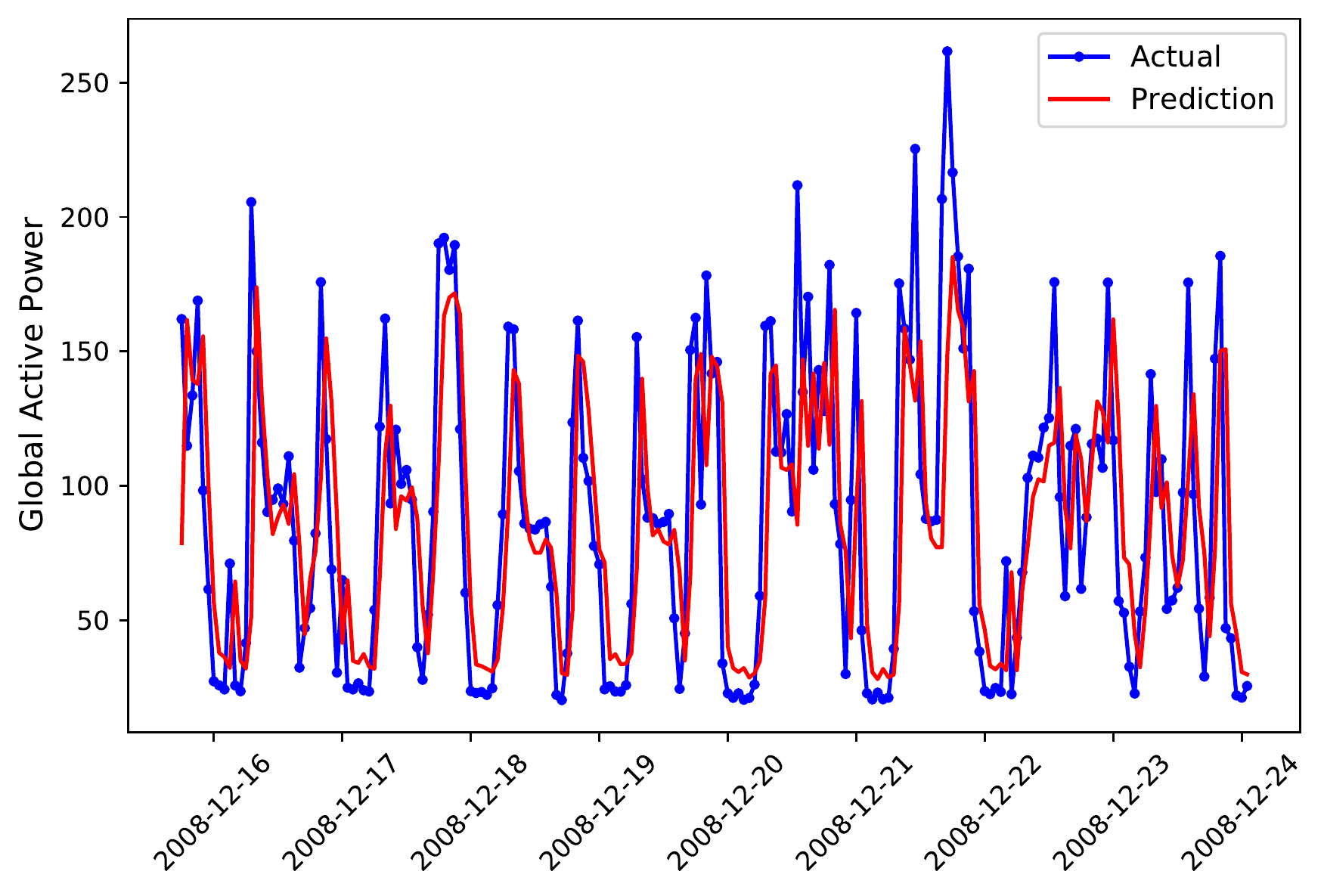}\label{fig:actual_vs_prediction}}
  \caption{Prediction Model Performance}
\end{figure}

\textcolor{black}{Model loss indicates how weak the model's predictions are. A loss close to zero means that the model's prediction is perfect and the model's performance is good.}
The proposed model loss started with a training loss of 0.0125 and a testing loss of 0.0079. Both of them continued to decline, reaching a training loss of 0.0095 and a testing loss of 0.0078. Fig. \ref{fig:loss} illustrates the loss performance over epochs.


Fig. \ref{fig:actual_vs_prediction} shows a snapshot of the model's perdition results for the global active power and the actual global active power values of 200 samples. As shown, the prediction follows the actual values pattern, while the error rate is low.

\subsubsection{\textcolor{black}{Energy} Saving}
\begin{figure}[!ht]
    \includegraphics[width=0.95\linewidth]{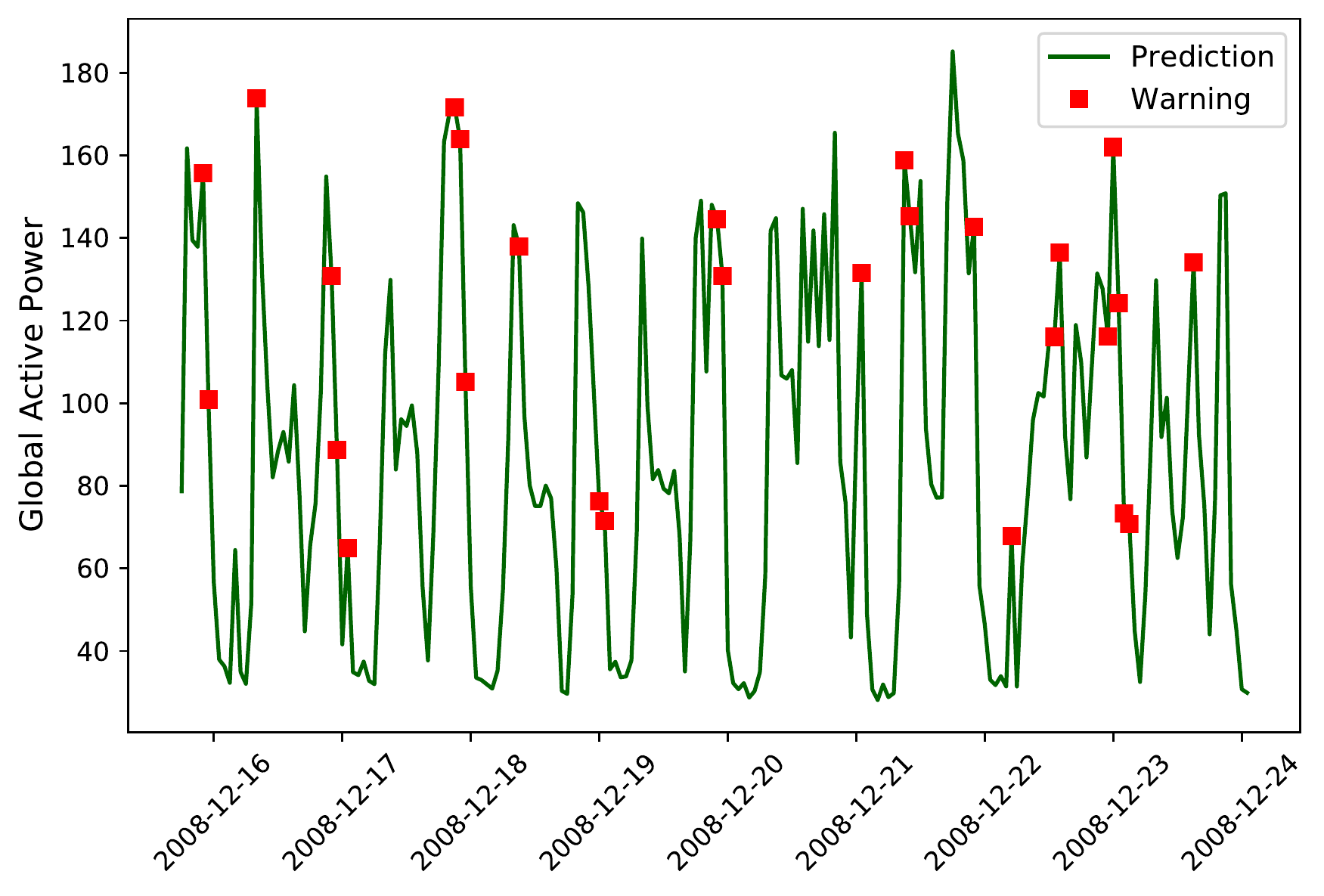}
    \caption{System Warnings}
    \label{fig:warning}
\end{figure}

Since the proposed system predicts the global active power for the next hour and controls the amount of used power to a certain threshold, a certain amount of \textcolor{black}{energy} will be saved. The system will control the power if the prediction is above the average consumption of the same hour of the same month and the same weekday for the past four years. \textcolor{black}{If the prediction is lower than the average it will not trigger the next action.}
The system has been tested on 200 samples from the testing data to know the amount of saved \textcolor{black}{energy}. As shown in Fig. \ref{fig:warning}, the system will issue warnings (red squares) on 27 hours from 200 predicted hours where the consumption is above the \textcolor{black}{historical average power consumption}. The total saved power amount for the 27 hours is 522.2 kilowatts.    

\subsubsection{Cost Saving}
As the system is tested on 200 samples and saved 522.2 kilowatts of global active power, the power cost will also be saved. In France, the price of electricity in kilowatt-hours for households is 0.182 euros. Therefore, the system will save €95.04 for a period of 200 hours.

\subsubsection{Global Intensity Reduction}
The energy intensity indicates the quantity of energy required per unit output or activity to produce a product or service. A high intensity indicates high energy used to provide a service, and low intensity indicates lower energy to provide a service. This affects the price or cost of energy conversion as low energy intensity indicates a low cost of energy conversion into the gross domestic product (GDP).
In the dataset, the global intensity is positively correlated with the global active power by 99\% so they both move in tandem, which means if the global active power decreases, the global intensity will also decrease.\\

To sum up, the proposed system in the use case scenario managed to reduce 522.2 kilowatts of \textcolor{black}{house} global active power over a 200-hours period, while helping to reduce €95.04 from wasted power cost. Moreover, the reduction of the global active power affected the global intensity positively, which will benefit the community as well. 

\section{Future Directions}
In the future direction, there are various open issues and areas of potential development that need to be investigated and tested:

Incorporating a user feedback subsystem into the proposed system is an important area for improving the learning and prediction of the AI model and the overall system. It is also important to measure the user satisfaction factor. This will help improve the experience, fix emerging issues, and keep the user more engaged.

Another potential development area is building a distributed version of the proposed system. The aim is to compare the distributed version to the current centralized version and test its impact on both technical and user experience aspects. Such as its effect on network efficiency, cost, data storing process, and system failures. 

Moreover, sending out instructive notifications to users to teach them about \textcolor{black}{the optimal} behavior is also a considerable approach in product and system design. \textcolor{black}{Users' responses} and use of instructive notices will speed up the process of changing behavior and make it easier, clearer, and more entrenched.

Furthermore, emphasizing the use of different AI/ML learning techniques is another development area to reduce the complexity of the AI model and increase its accuracy.

Finally, integrating technologies that add more control over security and data privacy is significant. 
The application of different security schemes, techniques, and technologies for different levels and processes of the system from data transmission to its storage and use will add more value to the system.

\section{Conclusion}
Since Internet-connected devices are increasingly being used; tracking, understanding, and influencing users' behavior to achieve material and moral profits has become a reality. This paper is the first attempt to investigate the concept of IoB, its workflow, benefits, challenges, and current industrial directions. It also proposes a trusted and comprehensible tracking-analyzing-influencing behavior system using IoB and XAI. Moreover, a use case of an electrical power consumption system that works for benefit of individuals, power companies, and society to save \textcolor{black}{energy} waste and cost. The collected results demonstrated that the system was able to save 522.2 kW of wasted power and €95.04 of its cost. Furthermore, the positive correlation between active power and power intensity will reduce the latter if the former is reduced.

 
\bibliographystyle{IEEEtran}
\bibliography{References}

\vspace{-1cm}
\begin{IEEEbiography}[{\includegraphics[width=1in,height=1.25in,clip,keepaspectratio]{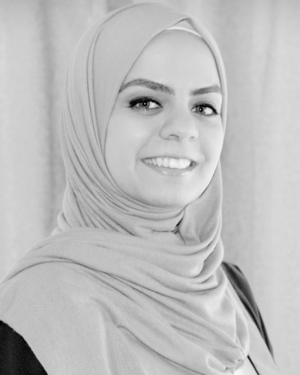}}]{Haya Elayan (M'20)} received the B.S. degree in computer information systems from The University of Jordan, Amman, Jordan, in 2018. She is currently pursuing the MRes degree in data science and analytics for health with the University of Leeds, Leeds, U.K. She worked as a Data Scientist in several companies. She applied machine learning techniques for data analytics, machine intelligence, feature selection, and recommendation engines, along with building and optimizing classifiers for product and operational capabilities. She also worked as a Researcher and a Data Scientist, where she was responsible for publishing academic papers and developing business insights.
\end{IEEEbiography}
\vspace{-1cm}
\begin{IEEEbiography}[{\includegraphics[width=1in,height=1.25in,clip,keepaspectratio]{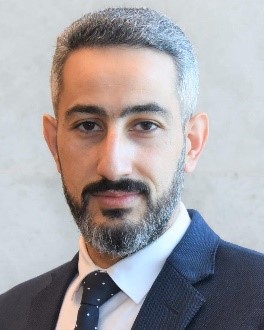}}]{Moayad Aloqaily (S'12, M'17)} received the Ph.D. degree in electrical and computer engineering from the University of Ottawa, Ottawa, ON, Canada, in 2016. He was an Instructor with the SYSC Department, Carleton University, Ottawa, in 2017. From 2018 to 2019, he was an Assistant Professor with the American University of the Middle East (AUM), Kuwait. From 2019 to 2021, he was the Cybersecurity Program Director and an Assistant Professor with the Faculty of Engineering, Al Ain University, United Arab Emirates. He has been the Managing Director of xAnalytics Inc., Ottawa, since 2019. He is currently with the Machine Learning Department, Mohamed Bin Zayed University of Artificial Intelligence (MBZUAI), United Arab Emirates. His current research interests include the applications of AI and ML, connected and autonomous vehicles, blockchain solutions, and sustainable energy and data management. He is an ACM Member and a Professional Engineer Ontario (P.Eng.). He has chaired and co-chaired many IEEE conferences and workshops. He started his own Special Interest Group on blockchain and application as well as internet of unmanned aerial networks. He has served as a Guest Editor for many journals, including IEEE Wireless Communications Magazine, IEEE Network, and Computer Networks. He is an Associate Editor of Ad Hoc Networks, Cluster Computing, Security and Privacy, and IEEE ACCESS. He has also been appointed as the Co-Editor-in-Chief of IEEE COMMSOFT TC ELETTER in 2020.
\end{IEEEbiography}
\vspace{-1cm}
\begin{IEEEbiography}[{\includegraphics[width=1in,height=1.3in,clip,keepaspectratio]{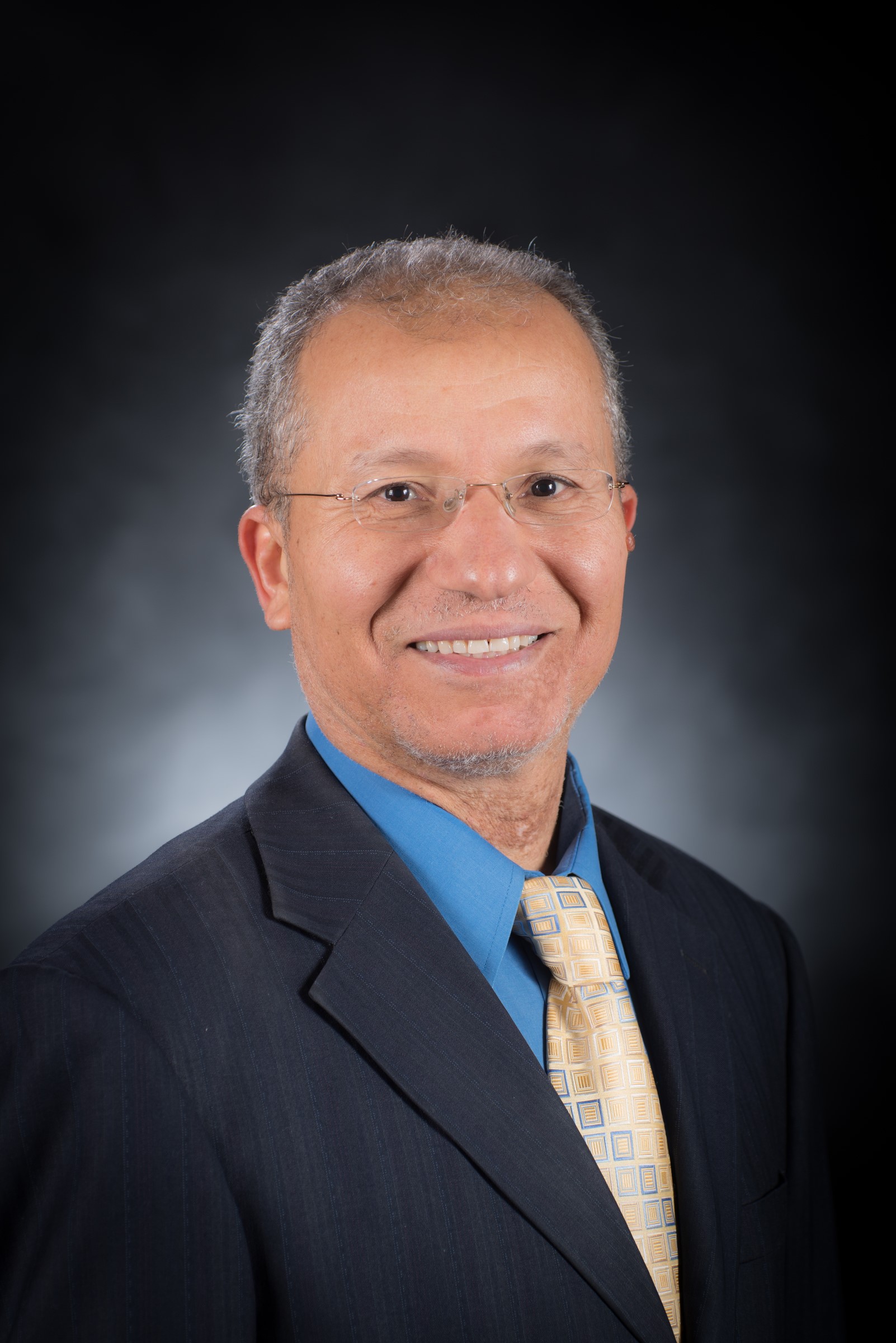}}]{Mohsen Guizani (S’85–M’89–SM’99–F’09)} received the BS (with distinction), MS and PhD degrees in Electrical and Computer engineering from Syracuse University, Syracuse, NY, USA. He is currently a Professor and the Associate Provost at Mohamed Bin Zayed University of Artificial Intelligence (MBZUAI), Abu Dhabi, UAE. Previously, he worked in different institutions in the USA. His research interests include applied machine learning and artificial intelligence, Internet of Things (IoT), intelligent systems, smart city, and cybersecurity. He has been an IEEE Fellow since 2009 and was listed as a Clarivate Analytics Highly Cited Researcher in Computer Science in 2019, 2020 and 2021. Dr. Guizani has won several research awards including the “2015 IEEE Communications Society Best Survey Paper Award” as well 4 Best Paper Awards from ICC and Globecom Conferences. He is the author of ten books and more than 800 publications. He is also the recipient of the 2017 IEEE Communications Society Wireless Technical Committee (WTC) Recognition Award, the 2018 AdHoc Technical Committee Recognition Award, and the 2019 IEEE Communications and Information Security Technical Recognition (CISTC) Award. He served as the Editor-in-Chief of IEEE Network and is currently serving on the Editorial Boards of many IEEE Transactions and Magazines. He was the Chair of the IEEE Communications Society Wireless Technical Committee and the Chair of the TAOS Technical Committee. He served as the IEEE Computer Society Distinguished Speaker and is currently the IEEE ComSoc Distinguished Lecturer. \end{IEEEbiography}
\begin{IEEEbiography}[{\includegraphics[width=1in,height=1.3in,clip,keepaspectratio]{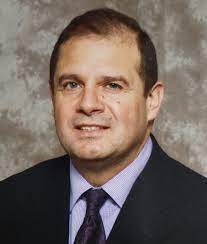}}]{Fakhri Karray}  received the Ph.D. degree from the University of Illinois at Urbana-Champaign, Urbana, IL, USA. He is currently the University Research
Chair Professor in electrical and computer engineering with the University of Waterloo, Waterloo, ON, Canada and the Co-Director of the University’s Institute for Artificial Intelligence. His research interests include intelligent systems design, big data analytic, soft computing, sensor fusion, and context-aware machines with applications to intelligent transportation systems, Internet of Things, and cognitive robotics. He has authored in these areas and has disseminated his work in journals, conference proceedings, and textbooks. He is a fellow of the Canadian Academy of Engineering and the Engineering Institute of Canada. He was an Associate Editor/Guest Editor for various journals, including the IEEE Transactions on Cybernetics, IEEE Transactions on Neural Networks and Learning, IEEE Transactions on Mechatrocnics, and IEEE Computational Intelligence Magazine.
\end{IEEEbiography}
\end{document}